\documentclass[letterpaper, 10 pt, conference]{ieeeconf}  %

\usepackage{array}
\newcolumntype{P}[1]{>{\centering\arraybackslash}p{#1}}
\usepackage{cite}
\usepackage{amsmath,amssymb,amsfonts}
\usepackage{graphicx}
\usepackage{textcomp}
\usepackage{algorithm}
\usepackage[noend]{algpseudocode}
\usepackage{mathtools}
\usepackage{svg}
\DeclarePairedDelimiter\norm{\lVert}{\rVert}%

\title{NV-Fogstore : Device-aware hybrid caching in fog computing environments}

\author{Khushal Sethi, Manan Suri \\
Department of Electrical Engineering\\
Indian Institute of Technology, Delhi \\
{\tt\small ee1160556@iitd.ac.in,  manansuri@ee.iitd.ac.in}}

\begin{document}

\maketitle
\thispagestyle{empty}
\pagestyle{empty}

\begin{abstract}
Edge caching via the placement of distributed storages throughout the network is a promising solution to reduce latency and network costs of content delivery. With the advent of the upcoming 5G future, billions of F-RAN (Fog-Radio Access Network) nodes will created and used for for the purpose of Edge Caching. Hence, the total amount of memory deployed at the edge is expected to increase 100 times.

Currently, used DRAM-based caches in CDN (Content Delivery Networks) are extremely power hungry and costly. Our purpose is to reduce the cost of ownership and recurring costs (of power consumption) in an F-RAN node while maintaining Quality of Service.

For our purpose, we propose $NV-FogStore$, a scalable hybrid key-value storage architecture for the utilization of Non-Volatile Memories (such as RRAM, MRAM, Intel Optane) in Edge Cache. 

We further describe in detail a novel, hierarchical, write-damage, size and frequency aware content caching policy $H-GREEDY$ for our architecture. 

We show that our policy can be tuned as per performance objectives, to lower the power, energy consumption and total cost over an only DRAM-based system for only a relatively smaller trade-off in the average access latency.

\end{abstract}

\section{Introduction}
\subsection{Rising Need for Edge Caching}
Edge caching via the placement of distributed storages throughout the network is a promising solution to reduce latency and network costs of content delivery. \cite{paschos2018role,mao2017survey}.
Employing caching in upcoming 5G wireless networks has been discussed and proposed in 3GPP 5G guidelines. For example, T-DOC R3-160688 \cite{gpp} proposes to place an edge cache at an LTE base station either embedded in eNodeB or standalone. \\
The objective of Local Caching (by memory embedded in the eNodeB-LTE Base Station), is to serve local user requests in a Fog-RAN Cell and to reduce network congestion in the backhaul and delivery latency. It can reduce redundancy of streaming popular multimedia contents, reduce duplicate mobile caching and prefetch preference learning-based predicted-content for users. \\
Currently deployed Content Delivery Networks (or CDNs) serve trillions of user requests a day from millions of nodes all across the globe and carry a majority of the internet traffic \cite{nygren2010akamai}. DRAM-based caches have been typically used in currently prevalent Content Delivery Networks \cite{nygren2010akamai,dilley2002globally}. \\
In the near future, billions of F-RAN (Fog-Radio Access Network) nodes will also be deployed for the purpose of caching. Hence, the total amount of memory deployed at the edge is expected to increase 100 times \cite{paschos2018role}.
\subsection{Why NVM based ?}
Emerging Non-Volatile Main Memory such as RRAM, MRAM, off-the-shelf Intel's Optane is useful for applications that need a large memory or require lower power and energy consumption. NVMs can store a large amount of data frequently to meet the constraints of low-latency and high bandwidth in 5G.
Further, the cost of ownership can be reduced by using NVMs. Compared to Flash, an NVM device offers 10$\times$ faster reads and has 5$\times$ better durability. By adding NVM to mitigate the access latency on the Flash/HDD layer, the overall cache size can be increased significantly. 

DRAM is up to 8x more expensive and uses 25x more power per bit than NVM \cite{eisenman2018reducing}. 

Our objective is to significantly reduce the total ownership cost by reducing the DRAM footprint and replacing it with NVM. If NVM is directly used as a DRAM-replacement without modification, it will wear out too quickly, due to its write durability constraints \cite{eisenman2018reducing}. A possible alternative is to use a Hybrid DRAM-NVM based system, with intelligent caching policies such that most of the writes are performed in DRAM and block reads are performed in NVM. 

\begin{figure}[!thpb]
    \centering
    \includegraphics[scale=0.5]{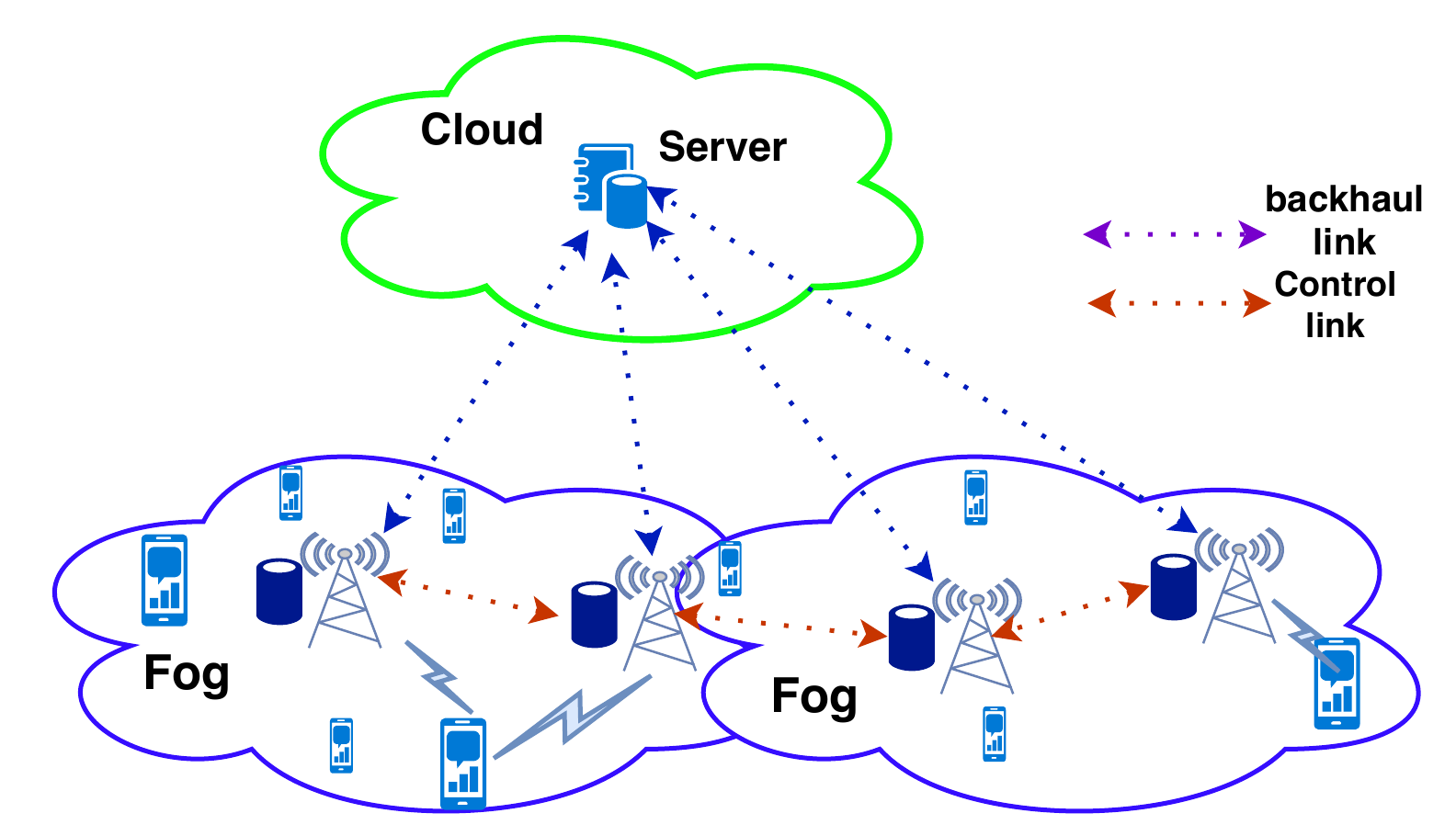}
    \caption{An Illustration of Fog-RAN which consists of Base-Station and mobile servers. Edge-Caching is used to reduce backhaul traffic.}
    \label{fig:my_label}
\end{figure}

\subsection{Contributions of this Work}
\begin{itemize}
      \item We propose a scalable hybrid key-value storage architecture for caching in fog computing environments.
      \item We formulate $H$-GREEDY, a hierarchical, write-damage, size and frequency aware content eviction/placement policy for our architecture.
      \item We compare our caching policy with traditional caching techniques by conducting extensive experiments on the traces of workloads envisioned in 5G for different configurations of file size, variance and Zipf's popularity.
      \item We show that our policy can be tuned as per performance objectives, to lower the energy consumption and cost over an only DRAM-based scenario for a relatively smaller trade-off in latency.
      \end{itemize}
      
\section{Background}
\subsection{Key-Value Store}
Key-value stores are used to persist data, enhance both read and write performance. Key-value storage is different from traditional storage in the implementation of its data access techniques. In a traditional computer, an OS page (typically of 4K Bytes is fetched), whenever a data-access is required. The pages are fetched continuously by looking up the B+-Tree file-system. Hence, the caching policies run on the OS page of a fixed size. 

In a key-value store, the path to retrieve data in a key-value store is a direct request to the object in memory or on disk. Hence, all the data is fetched simultaneously, as a block at the maximum bandwidth of the memory. For our architecture, we assume a simple exact-match hash table: which interfaces with common REST HTTP requests such as GET (retrieve the value), PUT (store a value for a given key) and DELETE (delete the key-value pair from store).

\subsection{Edge Caching Policies}

Web-traffic typically follows patterns in the popularity of requests, size of the content requested etc. This is usually modelled by the Independent Reference Model (IRM), which we briefly describe in subsection 1. 

Previous works \cite{jelenkovic2004optimizing,neglia2018cache} on finding the caching policy, has shown that they can be formulated as a Linear Utility Maximization problem. We briefly describe the background on how to  write the Knapsack problem for a single memory system in subsection 2.  Further it is described how an optimal caching policy is achieved from the solution of this Knapsack Problem. The notations used in the equations are given in Table I.

\subsubsection{Independent Reference Model}
The distribution of file size is assumed to be in a normal distribution of $ N(\mu,\sigma^2)$ where $\mu$ is the mean file size and $\sigma$ is the standard deviation, which can be determined statistically by trace generation over a period of time. The read density is given by the joint probability mass function of a file with parameters (size, popularity index) given as (s,z), \begin{equation}
 f_{size}=\frac{1}{\sqrt(2\pi)\sigma}  e^{\frac{(s-\mu)^2}{2\sigma^2}}
\end{equation}\\
The popularity of files for both read, update/delete operations are in a Zipfian distribution, independent of each other, with 80\% read requests and 20\% write requests. \begin{equation}
 n_{Z}(z)=\frac{1}{z^{\alpha}}
\end{equation}
where z is the file index and $\alpha \in [0.65,0.9]$. \\
Requests for content arrive according to the Poisson process with a rate corresponding to their popularity and mean access rate, the Poisson processes for different contents are independent \cite{fricker2012versatile}.  

\begin{table}[!htbp]
  \centering
  \caption{Summary of notations}
    \begin{tabular}{|P{0.8cm}|P{6.4cm}|}
        \hline $\mathcal{N}$ & Total Catalogue of Requests \\
    \hline $p_i$  & Estimated Popularity of Content $c_i$ \\
    \hline $w_i$  & Number of writes to Content i\\
    \hline  $s_i$  & Size of Content i \\
    \hline  $\beta$  & Number of Writes Exponent \\
  \hline $\mu,\sigma$ & Average File Size and Variance \\
 \hline $c_A,c_B$ & Content stored in DRAM, NVM \\
    \hline  $s_A, s_B$ & Size of DRAM, NVM \\
     \hline $r_A, r_B$ & Current Minimum Rank in DRAM, NVM \\
    \hline $t_A, t_B$ & Current Threshold for DRAM, NVM \\
     \hline $k_A, k_B$ & Rank for Storage in DRAM, NVM of Content k \\

    \hline
    \end{tabular}%
  \label{tab:precisioni}%
\end{table}%

\subsubsection{Caching Policy Derivation as a KnapSack Problem}

Cache policies have often been designed with the purpose to maximize the hit rate. The objective to maximize the hit-rate, can be reiterated as the contents that should be duplicated in the memory from the storage. This has been  formulated as a Knapsack Problem that describes the caching policies \cite{neglia2018cache}, as follows : 

\begin{equation}
\max\limits_{M\subseteq N} i\in M, \sum p_ic_i 
\end{equation}
\begin{center}
subject to $i\in M, \sum s_i \leq \phi$
\end{center}

It has been shown that  that under Zipf's law for popularities \cite{jelenkovic2004optimizing}, asymptotic hit ratio is optimized if we do a greedy caching of contents according to the ratio $c_i/s_i$.

Formally, an optimal eviction policy evicts a file $v$ in that satisfies the following:
\begin{equation}
v= \text{argmin} \{p_ic_i/s_i\} \ \forall i \in  \mathcal{N}
\end{equation}

\section{Related Work}
Several sizes, frequency, access-time and device-aware caching policies for edge caching have also been proposed in the literature \cite{shukla2016optimal,Neglia:2017:ACA:3133236.3149001,neglia2018cache,jelenkovic2004optimizing,shukla2017proactive,dabirmoghaddam2014understanding}. 

Writing files in a cache using the longest retention time damages the memory device thus reducing its lifetime.
However, writing using a small retention time can increase the content retrieval delay, since, at the time a file is requested, the file may already have been expired from the memory.

This motivates us to consider a joint optimization wherein we obtain optimal policies for jointly minimizing the content
retrieval delay (which is a network-centric objective) and the flash damage (which is a device-centric objective). Caching decisions now not only involve what to cache but also for how
long to cache each file. We design provably optimal policies and numerically compare them against prior policies.

Many of these approaches are shown to be combined in a linear utility maximization objective, which translates ot a Knapsack formulation for approaches that assume a single memory-based system \cite{neglia2018cache}. 

Reducing DRAM footprint has been explored in cloud-based scenarios, by utilization of NVM \cite{eisenman2018reducing}. Several key-value stores using DRAM/NVM based systems in cloud databases have also been investigated. \cite{liu2017librekv,bailey2013exploring,atikoglu2012workload}.

Hybrid NVM-DRAM Systems in Computing have been explored which involves techniques for hot-page migration and mitigation write amplification for management of Last-level caches, Persistent memory and Storage-class Memory (SCM). \cite{liu2017memos,Liu:2017:HCC:3079079.3079089,mittal2016survey}. 

However, no architecture and caching-policies are available for hybrid systems which take into account several device-aware properties of the system (such as endurance and unequal access times) for the cached-contents.

\section{NV-FogStore}
NV-FogStore targets the hybrid DRAM and NVM memory architecture, leveraging the NVM as the eventual persistent memory medium. 
Our Architecture (shown in Fig 2) has a flat-addressability of DRAM-NVM and content migration within the memories specified by the Memory Controller Instructions. The Tertiary Storage is assumed to be high endurance. 

The key features of our architecture are described below:

\textbf{Dynamic Memory Bank/Channel Rebalancing :} 
NV-FogStore uses bank partitioning to handle a faster request rate at a similar memory size. Since, in both DRAM and NVM, balanced bank parallelism is crucial in achieving desirable performance, the bank utilization is balanced during the allocation of banks to incoming contents. Memory requests across channels can also be balanced horizontally to achieve higher memory utilization. The admission policies and eviction policies (described in Section 4) take into account channel-utilization balancing using threshold ranking. The admission and eviction rank-threshold is tuned accordingly to the channel utilization and content migration takes place between NVM-DRAM channels to achieve better overall performance. \\
\textbf{Write Amplification Aware :} The NVM device in its lifetime can accommodate only a fixed total number of writes. Under a naive caching policy, the number of writes would significantly exceed the endurance limit of the device (i.e., its DWPD limit), which would cause it to rapidly wear out \cite{eisenman2018reducing}. Therefore, our policy should only stores blocks in NVM that have less frequent writes. 
 \begin{figure}[!tpbh]
    \centering
    \includegraphics[scale=0.55]{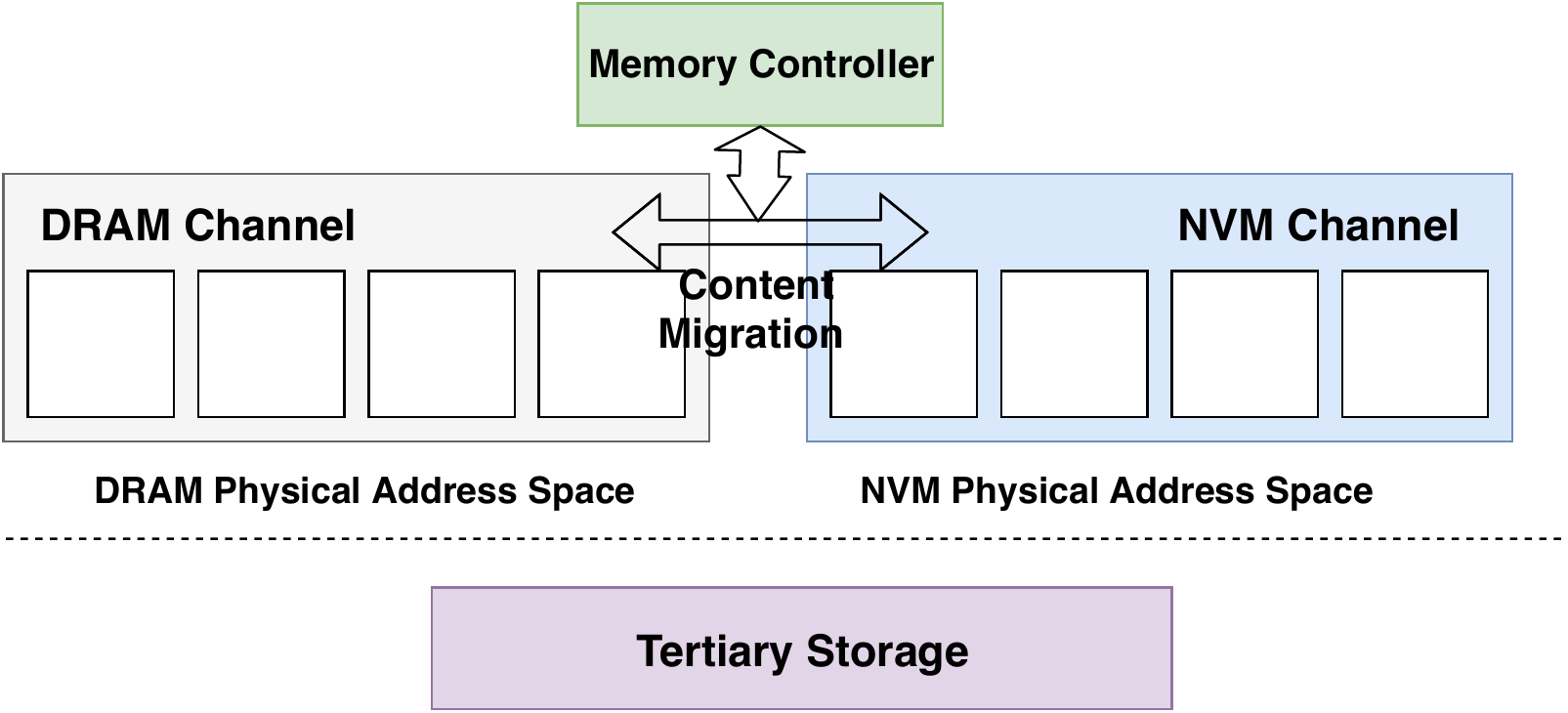}
    \caption{Flat Addressable Memory Architecture of NV-FogStore}
    \label{fig:my_label}
\end{figure}

\section{H-GREEDY Policy}
\subsection{Knapsack Formulation for a hybrid system}

Let  M and M* be set of contents that should be duplicated in the DRAM and NVM respectively in order to reduce the expected HDD workload generated from the next request. Then the caching policy can be written for a hybrid system in the utility maximization formulation as a simple extension of a single-memory system described above, as follows: \\

\begin{equation}
\max\limits_{M+M^*\subseteq N} i\in M,j\in M^* \sum p_ic_i+p_jc_j 
\end{equation}
\begin{center}
subject to $i\in M, \sum s_i \leq \phi, j \in M^*, \sum s_j \leq \Phi$   
\end{center}

\subsection{Hierarchical Rank-based Policy}

From the insights of the knapsack formulation above, we  propose a novel generalized  algorithm, H-GREEDY, (Algorithm  1)  for caching contents in NV-FogStore and dynamically  moving  data  between  the  NVM  and  DRAM memories.

For our hierarchical greedy method, every content/file is provided with two ranks, each for content storage in DRAM and NVM. 

Rank for content storage in DRAM (denoted by $k_A$) is given by, $(p_{k} w_{k}^h/s_k)$. Since minimum rank contents will be removed from memory, this rank means that content with a high number of reads and writes and lower file size will not be evicted to NVM. 

Rank for content storage in NVM (denoted by $k_B$)  is given by, $(p_{k} /s_k w_{k}^h)$. This rank indicates that content with lower popularity, higher file size and a high number of writes will be evicted from NVM to HDD Storage. 

The popularity/read and write rate for content is updated only at the arrival of a request for that content. All the estimates are also updated every $\tau$ cycles, to do a complete refresh of the statistics stored. Due to the noisy popularity estimates, randomness is a collateral effect in our scheme.

\subsection{Convergence Issues}
Running caching policies is difficult due to various convergence issues \cite{jiang2017lru}. Fast convergence is paramount to the performance of these policies in real-world scenarios. To allow this, we also define the threshold number for each DRAM and NVM  ($\alpha, \beta$), which is dependent on the utilization of the memories (Algorithm 1). 

We also track the minimum rank content currently stored in DRAM and NVM ($r_A, r_B$). The combination of these two numbers, threshold and minimum rank, is used to make the rank-threshold product ($\alpha*r_A$, $\beta*r_B$) will be used for our admission and eviction policies. This allows better convergence at a lower utilization of memories since early requests for storing content are not rejected because the threshold number is also lower. 

\subsection{Overall Execution}

The overall execution of the H-GREEDY policy is in the following manner: Every incoming object request is added to a queue. If the object is already stored in the system, its properties are updated, otherwise, it allocated a store (DRAM/NVM) or it is not stored, depending upon the admission policy. The store allocation for the content is done is by the rank-threshold product described above. Hence, for under-balanced channel allocation, the utilization of the corresponding channel can be increased by the tuning of the threshold parameter. Further, it is also allocated a bank depending upon where it is stored.

The threshold for both DRAM/NVM is then updated based on the current utilized capacity of the memories.

After a certain $\Delta$ requests, the eviction policy is run on the contents with updated properties, content evictions carried out from the DRAM to NVM (and vice-versa), or deletion of the content from DRAM or NVM. Here, $\Delta$ is a parameter, known during the execution of the policy and can be quickly determined by the request rate and the total number of banks available in the entire system. 
The data is evicted using the same threshold-rank described earlier. Hence, our eviction policy moves content in a two-level greedy approach that is write-damage aware.

At last, after a certain $\tau$ cycles of time, the properties of all the objects are reset, and cold objects are removed from the system. Here, $\tau$, is a certain multiple of the DRAM latency cycles determined and set by empirical observation.
\begin{algorithm}
\caption{H-Greedy}
\label{euclid}
\begin{algorithmic}[1]
\State \emph{\textbf{Procedure : H-Greedy}}
\For {each cycle}
\If {$object_i \in M+M^*$ }: 
\State Add $object$ Request to Queue 
\State Update $object_i$ Properties
\Else
\State $object.store \gets AllocateStore(object)$
\State $Allocate Bank(object)$
\EndIf
\State \textbf{Adjust DRAM and NVM Threshold}
\If {($utilization_{mem}<0.5$)} $thr_{mem} \gets 0$ 
\Else \ $thr_{mem} \gets utilization_{mem}$
\EndIf
 \State \emph{\textbf{Content usage predicted}}
 \State $RunEvictions(Update Object)$
\If{($total\_{cycles}\%\tau=0$)}:\\
 \ \ \ \ \ \ \ \ \ \  Reset $object.props$, Remove cold $objects$
\EndIf
\EndFor
\State \emph{\textbf{End Procedure}} 
\end{algorithmic}
\end{algorithm}

\begin{algorithm}
\caption{Eviction and Admission Policies}
\begin{algorithmic}[1]
\State \emph{\textbf{Procedure : RunEvictions}} 
\For {key in Updated:}
\State   $k_A, k_B$ $\gets$ GetRank(key)
\State \textbf{case key $\in$ DRAM} :
        \If{$(k_A<t_A*r_A$ and $k_B>t_B*r_B)$}: \\
 \ \ \ \ \  \ \ \ \ \ \ \ \ \ \ Evict to NVM
        \Else
            \If{($k_A<t_A*r_A$ and $k_B<t_B*r_B$)}: \\
 \ \ \ \ \  \ \ \ \ \ \ \ \ \ \ Evict to HDD/Delete
            \EndIf
        \EndIf
\State \textbf{case key $\in$ NVM} :
    \If {$(k_A>t_A*r_A$ and $k_B<t_B*r_B)$}:\\
\ \ \ \ \  \ \ \ \ \ \ \ \ \ \ Evict to DRAM
    \Else
            \If{$(k_B<t_B*r_B)$}: \\
\ \ \ \ \  \ \ \ \ \ \ \ \ \ \ Evict to HDD/Delete
            \EndIf
    \EndIf
\State \textbf{case key $\in$ HDD} :
\If{($k_A>t_A*r_A$ and DRAMFree($key_{size}$))}:\\
   \ \ \ \ \  \ \ \ \ \  \textbf{return} DRAM
\EndIf 
\If{$(k_B>t_B*r_B$ and NVMFree($key_{size}$))}:\\
  \ \ \ \ \  \ \ \ \ \   \textbf{return} NVM
\EndIf   
\EndFor
\State \emph{\textbf{End Procedure}} 
\end{algorithmic}
\begin{algorithmic}[1]
\State \emph{\textbf{Procedure : AllocateStore}} 
\State  $k_A, k_B \gets$ GetRank($key$)
\If{($k_A>\alpha*r_A$ and DRAMFree($key_{size}$))}:
    \If{($r_A>k_A$)}: $r_A=A$;
    \textbf{return} DRAM
    \EndIf
\EndIf
\If{$(k_B>\beta*r_B$ and NVMFree($key_{size}$))}:
    \If{($r_B>k_B$)}: $r_B=k_B$; \textbf{return} NVM
    \EndIf
\EndIf
\State \textbf{return} HDD
\State \emph{\textbf{End Procedure}} 
\end{algorithmic}
\begin{algorithmic}[1]
\State \emph{\textbf{Procedure : GetRank}}
\State \emph{\textbf{Rank of Content}}
\State $k_A \gets p_k/(w_k^h*s_k)$; $k_B \gets (p_k*w_k^h)/s_k$ 
\State \textbf{return} $k_A, k_B$
\State \emph{\textbf{End Procedure}} 
\end{algorithmic} 
\end{algorithm}

\section{Discussion and Performance Analysis}
\subsection{Experimental Setup}
We simulate the structure of requests using open-source web-traffic trace-generator \cite{berger2017adaptsize}, each file is given a size and popularity index with values sampled from the respective distributions. The Hybrid-Memory System is simulated in an in-house cycle-accurate simulator (Python-based adapted from \cite{stevens2013integrated}, modified with policies described), which runs on request traces generated. Our System configuration consists of 256GB DDR4 DRAM  and 4TB NVM configured with JEDEC DDR4-SDRAM \cite{ddr} and Intel Optane Standards respectively \cite{optane,liu2017memos}. Several traces are generated with varying mean file size, file size variance and Zipf's Parameter, each consists of $\sim$ 25 million user-requests and $\sim$ 1 million unique objects. 

\subsubsection{Trace Configurations}
For our evaluation in the sections below, we can test for different configurations of the NVM/DRAM System (defined by their different cache capacities). We run the system on 9 traces, that is for three different values of the Zipf's parameter = [0.9,0.8,0.7] and three different values of mean value size = [1MB, 5MB, 10 MB] for each of the values of the Zipf's parameter. For simplification purposes, we assume the file size variance to be equal to the mean file size. 

\subsubsection{Cache Capacity}

While our setup seems too dependent on the assumptions of the configuration described above, our evaluation is actually agnostic to the configurations. So, to explain this we define the term cache capacity, as the ratio of the size of the respective DRAM and NVM memories to the size of the total request trace. 

Hence, mathematically DRAM cache capacity can be written as $(\frac{\phi}{\norm{\mathcal{N}}*\mu}$) and NVM cache capacity as ( $\frac{\Phi}{\norm{\mathcal{N}}*\mu})$ (from the notations defined in Table I). 

In other words, cache capacity signifies what percentage of all the content requested can altogether be stored in the given system. Since the number of requests in the trace is fixed, and system configuration in size of memory is fixed, the cache capacity is only dependent on the mean file size. 

For the range of the mean file size described above (i.e. [1MB, 5MB, 10MB]), the cache capacity for the (DRAM, NVM) pairs can be written as : [(0.256,4), (0.05,0.8), (0.12,0.2)]. This means, the first pair given as (0.256,4), the DRAM can store 25.6\% of the total size of requested content trace and NVM can store 4 times (400\%) of the total size of the trace. Similarly, in pair 2, DRAM capacity is  5\% and the NVM capacity is 80\%, and pair 3, it is 1.2\% and 20\% respectively. This covers a significant range of real-world hybrid caching configurations.  

By defining the term caching capacity, we can symbolically evaluate our policy without the dependency on the configuration. 
\begin{table}[!htbp]
  \centering
  \caption{The parameters of NVM and DRAM}
    \begin{tabular}{|P{1.4cm}|P{5.4cm}|}
    \hline DRAM System \cite{ddr} & Read(Write) Energy=51.2nJ, Read (Write) Bandwidth=75 GB/s, Read(Write) Latency=75 ns, Standby Power=1W/GB, Endurance=Very High \\
    \hline NVM System \cite{optane,liu2017memos} & 
    Read(Write) Energy=102.4(512)nJ, Read (Write) Bandwidth=2.2 (2.1) GB/s, Read(Write) Latency=10$\mu$s, Standby Power=0.1W/GB, Endurance= 30DWPD\\
    \hline
    \end{tabular}%
  \label{tab:precisioni}%
\end{table}

\subsection{Performance Metrics}

The performance metrics considered in the evaluation of the system include the cost (initial purchase cost of the memory itself along with the replacement costs of memory due to wear) power-consumption of the system and an average latency of fetching the contents (time of response to user requests). These performance metrics can then be tracked, in our comparison of NV-FogStore with a DRAM-only architecture

Although, cost can be difficult to measure, (including replacement costs due to wear) as they may be vendor-specific, we track the metric of Cost Benefit Ratio (CBR, the higher the better), defined as follows :

\begin{equation}
CBR = \frac{Cost_{dram}(\phi)+ Cost_{nvm}(\Phi)*\frac{Lifetime_{nvm}}{Lifetime_{dram}}}{Cost_{dram}(\phi+\Phi)}
\end{equation}

This takes into account the different lifetimes and wear (predominantly due to writes) of DRAM and NVM in the overall cost. 
\begin{figure*}[!tpbh]
    \centering
    \includegraphics[scale=0.30]{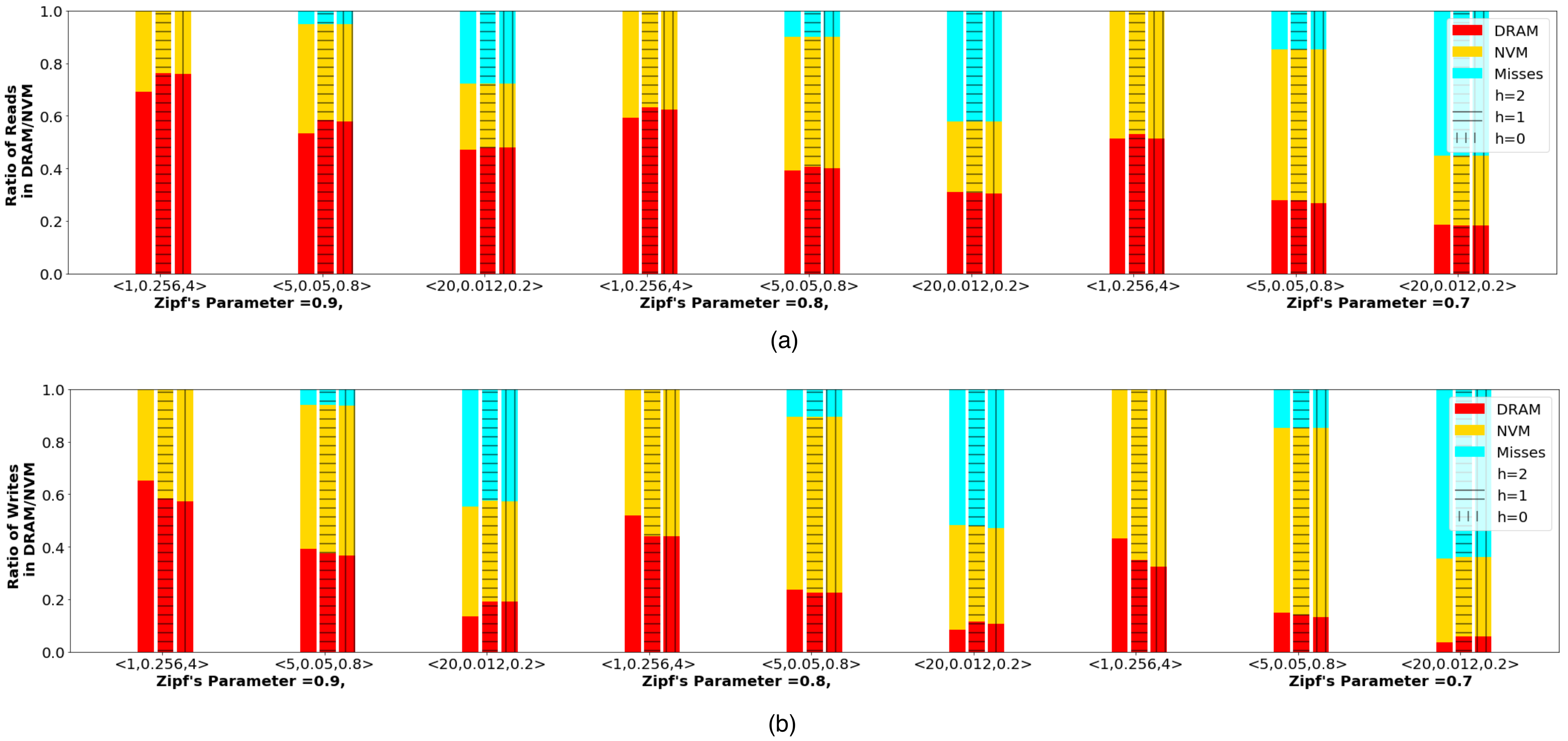}
    \caption{(a) Reads in DRAM/NVM/Misses(Served by Tertiary Storage), (b) Writes in DRAM/NVM/Misses(Served by Tertiary Storage) with varying parameter $h$, Zipf's Parameter and Mean File Size $(\mu)$. The X-axis Labels are $<$Mean File Size (in MB), DRAM Cache Capacity, NVM Cache Capacity$>$. Cache Capacity is defined as the ratio of Capacity of Memory to the Total Trace Size.}
    \label{fig:my_label}
\end{figure*}
\begin{figure*}[!tpbh]
    \centering
    \includegraphics[scale=0.20]{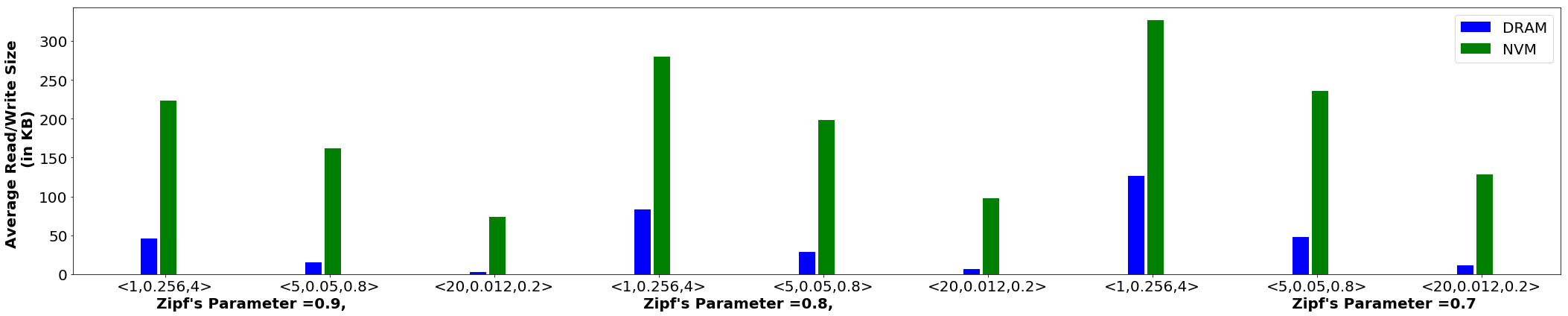}
    \caption{Average Size of Read/Write Request in DRAM/NVM, with varying Zipf's Parameter and Mean File Size $(\mu)$. The X-axis Labels are $<$Mean File Size (in MB), DRAM Cache Capacity, NVM Cache Capacity$>$. Cache Capacity is defined as the ratio of Capacity of Memory to the Total Trace Size.}
    \label{fig:my_label}
\end{figure*}
\subsection{Evaluation of the Hierarchical Greedy Policy}

\subsubsection{Write-Aware Evaluation}

We run our policy on the traces generated (as described in section A) with varying the parameter h in our device-dependent model. And we compare our policy with the previous state-of-the-art  approach \cite{neglia2018cache} in this subsection.
write-aware policy actually works,

Figure 3 (a) and (b) depicts the total number of reads and writes served in DRAM and NVM with varying the parameter $h$ (from 0 to 2), of our policy on the nine configurations of traces described in Section 5A.  
 
In each of the nine configurations, increasing the parameter h, the fraction of writes served by NVM reduces and the fraction of reads served by NVM increases. Since, the reads served will take more time, and writes are reduced (less wear), this variation demonstrates the access latency-endurance trade-off. Since 3GPP standards require requests to have a Normalized delivery time $(NDT)$ of less than 4 milliseconds, it is possible to store several files in NVM using appropriately tuned parameters in our policy, without worrying about the latency. 

For further analysis across the configurations, as the mean file size is increased (for a fixed Zipf's Parameter) (Figure 3), the DRAM/NVM cache capacity decreases. Hence, the number of misses increase. However, even at very low cache capacities, it is visible that most of the requests are served by the DRAM/NVM System.

Similarly, keeping the mean file size fixed, and analyzing across increasing Zipf's parameter (from the 0.7 to 0.9), it is seen more requests can be served by the DRAM, thus reducing the overall access time. This is because it is easier to estimate the popularity's of each file correctly, with a more skewed distribution. Also, it can be seen that most of the reads and writes can be served even when DRAM and NVM are a small fraction of required capacity, and the fraction of requests served increases on increasing the Zipf's parameter. For modern internet traffic, the parameter lies close to 1, so we can expect most of the requests to be served in a relatively short time, with very few misses. 

\subsubsection{Request Size-Aware Evaluation} 
Fig 4. shows the average size of the file stored in DRAM and NVM, for the nine trace configurations. As seen, the average size of the file stored in NVM is much larger, thus leveraging as a block device (for block reads) compared to DRAM. This proves the working of the size-aware nature of our policy.

Fig 4, can be analyzed for varying mean file size and Zipf's parameter. On decreasing the cache capacity, the average file decreases, because only the smallest most popular files are given priority. On increasing the Zipf's parameter, the distribution gets more skewed and the average file size that needs to be stored reduces. This can be reasoned as the content popularity's decrease faster due to more skewed distribution, size of the file in admission reduces for the same rank.

\subsection{Comparison with Optimal Offline Policy}
We perform a comparison of our methodology with the optimal offline policy. This is necessary to determine the effectiveness of our policy in real-time execution (where statistics of the data items, such as popularity's of the content are determined on the go and are noisy), compared to an offline version where the popularity's, number of writes of and the sequence of requests for each content are known beforehand. 

The offline optimal policy selects content on the H-Greedy method and allocates the store on the rank calculated from the known properties beforehand.

Figure 5 shows in detail the comparison, across all the nine configurations, when the policy is run with parameter $h = 2 $. 

For simplification, we discuss average statistics across the nine configurations here. The average number of reading requests served by DRAM/NVM/Misses are in ratio $48.8:39.2:11.8$ for the offline policy, whereas, in the online policy, they are in ratio $43.7:40.5:15.7$. The average number of write requests served by DRAM/NVM/Misses are in the ratio $51.33:22.5:26.1$ for the offline policy, whereas, in the online policy, they are in the ratio $51.62:28.92:19.44$.
So, in comparison with an optimal offline policy, our online policy does comparatively well in serving read requests, although the ratio of write requests served by NVM is 28\% in online policy is worse compared to 22\% in an optimal offline policy.

This makes sense because the number of read requests are larger and read popularity estimation is easier. Better heuristics may be created for closing the gap between offline and online policies.

\begin{figure*}[h]
    \centering
    \includegraphics[scale=0.30]{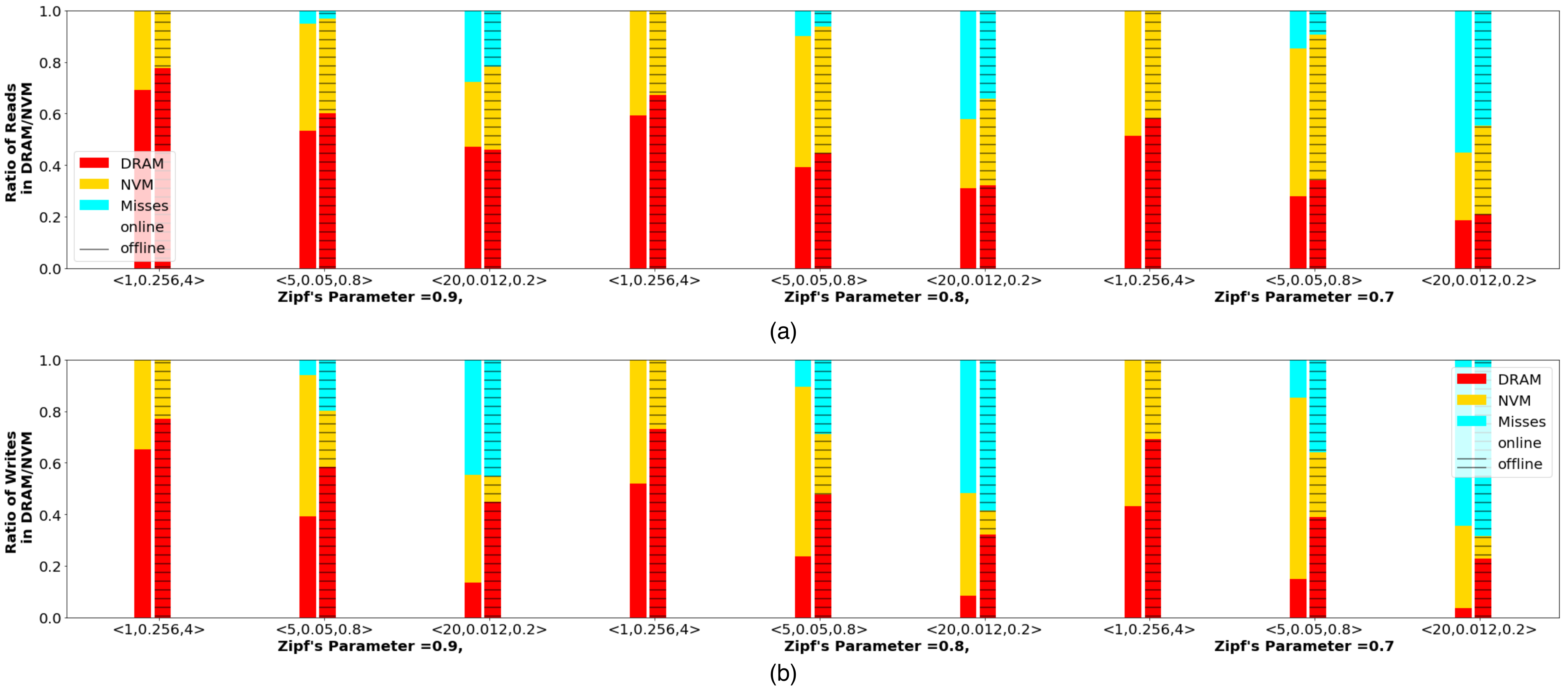}
    \caption{(a) Reads in DRAM/NVM/Misses(Served by Tertiary Storage), (b) Writes in DRAM/NVM/Misses(Served by Tertiary Storage) with $h$ = 2, Zipf's Parameter and Mean File Size $(\mu)$ showing comparison of Online and Offline Policies. The X-axis Labels are $<$Mean File Size (in MB), DRAM Cache Capacity, NVM Cache Capacity$>$. Cache Capacity is defined as the ratio of Capacity of Memory to the Total Trace Size.}
    \label{fig:my_label}
\end{figure*}

\subsection{Benchmarking with DRAM-Based System}
Figure 6 compares NV-FogStore with a DRAM-only architecture. The parameters of the NVM and DRAM System used for this comparison are given in Table 2. 

Since DRAM forms less than 10\% of the capacity in NV-FogStore, both power consumption and cost are dramatically decreased. Although, the average access time is increased compared to a DRAM-based system, it is about 2 $\times$ lower than an only NVM-based system. The access time can be increased because the effect on the overall delivery time is negligible by Fog-RAN standards \cite{gpp}. 

Even under different objectives, our proposed policy can be easily tuned to match the owner's requirement. Preventing the storage of content with a large number of writes in NVM to increase lifetime is another trade-off with the latency of content fetch as it means that more popular contents will now be stored in NVM instead of DRAM. 

Our scheme is configurable to explore the trade-off between the performance taken as average access latency and cost-benefit ratio.

\begin{figure}[h]
    \centering
    \includegraphics[scale=0.35]{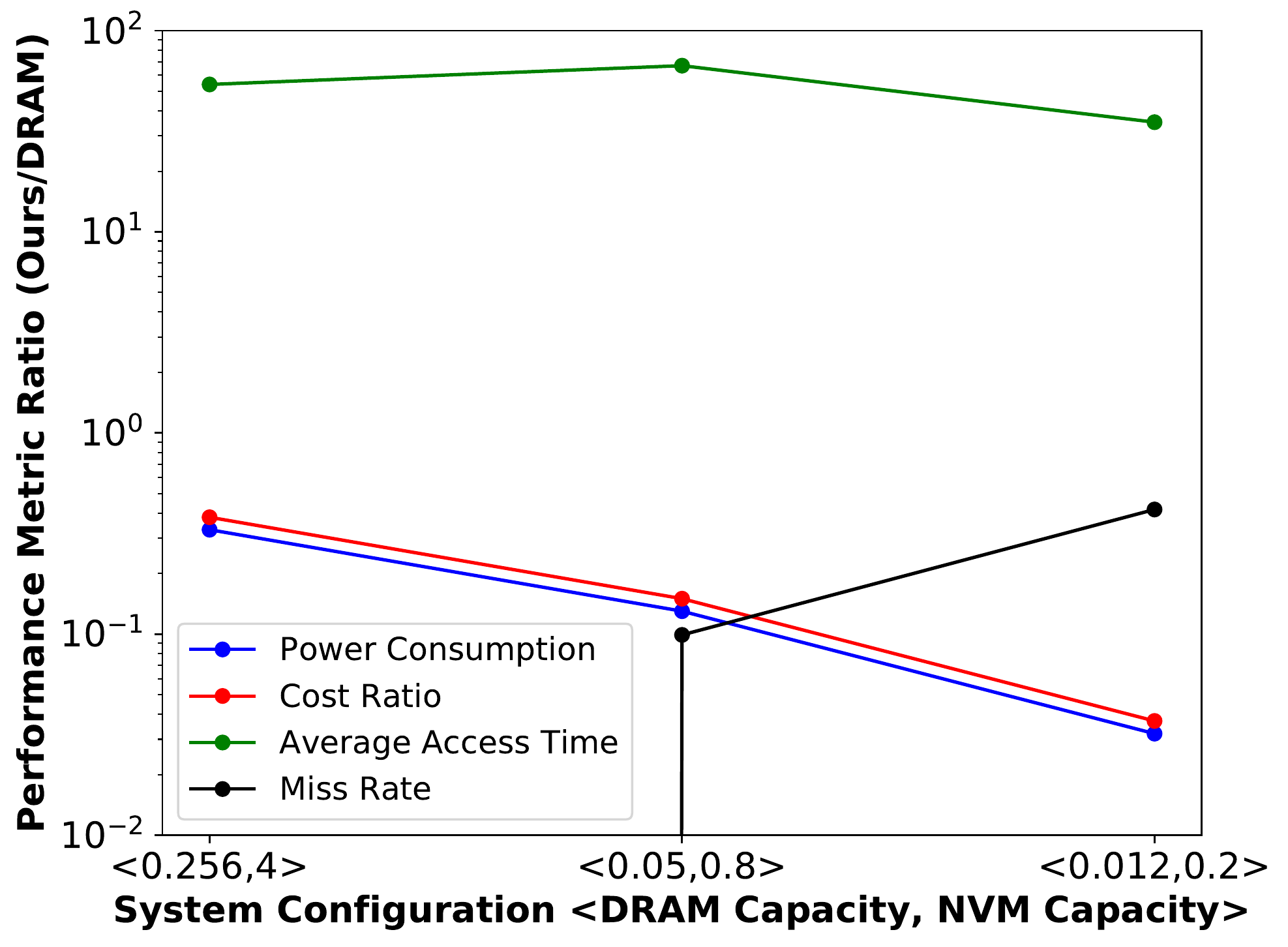}
    \caption{Effect of System Configuration of NV-FogStore on various Parameters such as Average Access Time, Power Consumption, Cost and Miss Rate. Miss Rate is zero for the first configuration.}
    \label{fig:my_label}
\end{figure}

\section{Conclusions}
In this paper, we investigate utilizing non-volatile-dram memory for content caching to optimize cost and power consumption in F-RAN (Fog-Radio Access Networks) and CDN (Content Delivery Networks) with storage under modern-5G workloads. An optimal caching policy $P$, which takes into consideration the user requests, user preference learning-based recommendation files, local popularity of content to determine the file trace produced and it's key-value commands. Mechanisms of such a caching policy maybe hand-optimized \cite{qazi2019optimal,ren2019profitable, paschos2019learning}, incorporating a machine-learned based advice or value-function-based reinforcement learning techniques \cite{tao2019content,song2019one,sadeghi2018reinforcement,sadeghi2019adaptive}. \\
Our work is focused on the design of memory policies which meet the demands of these emerging workloads while being deployed at the edge and give a significant reduction in energy costs. For this, we proposed a write-amplification aware, \textbf {hierarchical greedy} caching policy which is aware in both the size and frequency of request objects. Further optimizations are done, based on threshold-rank product to allow optimum convergence. The performance of our policies and are compared against prior policies using simulations.
To the best of our knowledge, this is the first study exploring the usage of NVM in a Fog-RAN environment. Future work may involve the exploration of Probabilistic caching approaches based on Dynq-LRU or Simulated Annealing based approaches for Hybrid Systems.

\bibliographystyle{IEEEtran}
\bibliography{ref}

\end{document}